\documentclass[prl,showpacs,twocolumn,floatfix,superscriptaddress]{revtex4}
\usepackage{epsfig}
\usepackage{amsmath,amssymb,mathrsfs}

\begin{document}

\title{Fano-Rashba effect in a quantum wire}
\author{David S\'anchez}
\affiliation{Departament de F\'{\i}sica, Universitat de les Illes Balears,
E-07122  Palma de Mallorca, Spain}
\author{Lloren\c{c} Serra}
\affiliation{Departament de F\'{\i}sica, Universitat de les Illes Balears,
E-07122  Palma de Mallorca, Spain}
\affiliation{Institut Mediterrani d'Estudis Avan\c{c}ats IMEDEA (CSIC-UIB), 
E-07122 Palma de Mallorca, Spain}

\date{14 March 2006}

\begin{abstract}
We predict the occurrence of Fano lineshapes in a
semiconductor quantum wire
with local spin-orbit (Rashba) coupling.
We show that the Rashba interaction acts in a strictly
one dimensional channel as an
attractive impurity, leading to the formation of purely bound states. 
In a quasi-one dimensional system these bound states
couple to the conduction ones through the Rashba intersubband mixing,
giving rise to pronounced dips in the linear conductance
plateaus. We give exact numerical results
and propose an approximate model capturing
the main ingredients of the effect.
\end{abstract}

\maketitle

In the Fano effect \cite{fano} the
interference of a bound state
immersed in a continuum with a nonresonant channel gives rise
to characteristic asymmetric lineshapes.
Fano lineshapes have been observed in
atomic physics \cite{ada49}, Raman scattering \cite{cer73} and
mesoscopic electron transport \cite{gor00,kob02}, just to mention a few.
In all these cases, the origin of the binding potential was clear
and the spin degree of freedom played a passive role.
Then, it is natural to ask whether a scattering potential coupling
the spin and orbital coordinates is able to
generate Fano lineshapes alone.  In this Letter we show that the
answer is yes.

The latter question is not only of academic interest.
There is a continuously growing interest in the potential applications
of spintronic devices, where the electron spin is the information
carrying basic unit \cite{zu04}.
Of special relevance is the experimentally
demonstrated manipulation of spin-orbit potentials in low dimensional
semiconductor systems via electric gates \cite{nit97}.
A type of spin-orbit coupling prominent in certain quantum well
heterostructures (typically, InAs based systems)
is the Rashba interaction \cite{ras60}.
It arises when the confinement potential which defines the quantum
well is a function of the direction perpendicular to the two-dimensional
(2D) electron gas (say, the $z$-direction).
This implies a structural inversion asymmetry which, importantly,
can be tuned with an external gate potential \cite{nit97}.
To lowest order in the momentum, the Rashba Hamiltonian reads,
\begin{equation}\label{eq_hr}
H_R=\frac{1}{2\hbar} (\{\alpha,p_y\}\sigma_x  - \{\alpha,p_x\}\sigma_y) \,,
\end{equation}
where the Rashba strength $\alpha$ is proportional to the electric field
along $z$, $\vec p=(p_x,p_y)$ is the 2D momentum operator
and $\sigma_i$ ($i=\{ x,y,z \}$) are the Pauli matrices.
The anticommutators in Eq.\ (\ref{eq_hr}) ensure a
hermitian Hamiltonian when $\alpha$ is nonuniform.
Attention to semiconductor spintronics has boomed partly due
to the Datta-Das proposal \cite{dat90}
of an all-electrical spin transistor
using the precession effect of a Rashba field onto an injected
stream of polarized electrons. The idea is based on a quasi-1D ballistic
channel attached to ferromagnetic contacts \cite{dat90}.
Hence, it is of fundamental importance
to investigate transport of electrons subject to spin-orbit coupling
along quantum wires \cite{sat99,mir01,gov02,scha04,zhan05}.
Furthermore, Egues {\it et al.}
suggest \cite{egu02} a quantum wire with local Rashba interaction
within a beam-splitter configuration as a tool for entanglement
detection via shot noise. 

Consider the strict 1D limit of a ballistic
quantum wire with local Rashba interaction $\alpha(x)$:
$\alpha(x)=\alpha_0$
(constant) for $0<x<\ell$ and zero elsewhere.
Then, the Hamiltonian in the effective mass approximation is
$H=p_x^2/2m-\{\alpha(x),p_x\} \sigma_y /2\hbar$
and the electron wavefunction
may be expanded, $\psi(x)=\psi_1(x)\chi_+ + \psi_2(x)\chi_-$,
in the spinor eigenstates $\chi_{\pm}$ of $\sigma_y$.
We now apply the following gauge transformation:
$\psi_{1,2}\to \psi_{1,2} \exp{(\pm i \int^x{k_R(x') dx'})}$,
where $k_R(x)=m \alpha(x)/\hbar^2$.
As a result, the Schr\"odinger equation becomes
$-(\hbar^2/2m)\psi_{1,2}''=[E+(\hbar^2 k_R^2/2m)]\psi_{1,2}$, i.e., 
the textbook problem of a square-well
potential of length $\ell$ and depth $-m\alpha_0^2/\hbar^2$.
This simple argument shows that 
a local Rashba interaction in the 1D limit forms bound states
for negative energies.
This important fact has been, to the best of our knowledge,
overlooked in many works, which consider only positive-energy states.
Exceptions are Ref.~\cite{val04}, which predicts bound states
in a Rashba disk within a 2D electron gas,
and Ref.~\cite{cso05}, which discusses the energy spectrum
of 2D Rashba billiards.

\begin{figure}
\centerline{
\epsfig{file=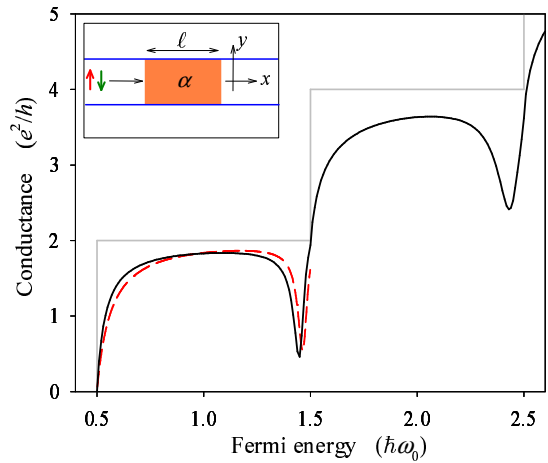,angle=0,width=0.40\textwidth,clip}
}
\caption{(Color online)
Linear conductance of an electron waveguide with
local Rashba interaction (solid line).
The solution given by the coupled-channel method is shown
with dashed line. We set Rashba strength $\tilde{\alpha}=1$
in units of $\hbar\omega_0 l_0$ and dot size
$\tilde{l}=0.75$ in units of $l_0$ with
$l_0=\sqrt{\hbar/m\omega_0}$
and $\hbar\omega_0$ the transversal
confinement length and energy scales, respectively.
Gray line: the case without Rashba interaction.
Inset: Sketch of a quasi-1D system with local Rashba
interaction.
}
\label{fig1}
\end{figure}

\begin{figure}
\epsfig{file=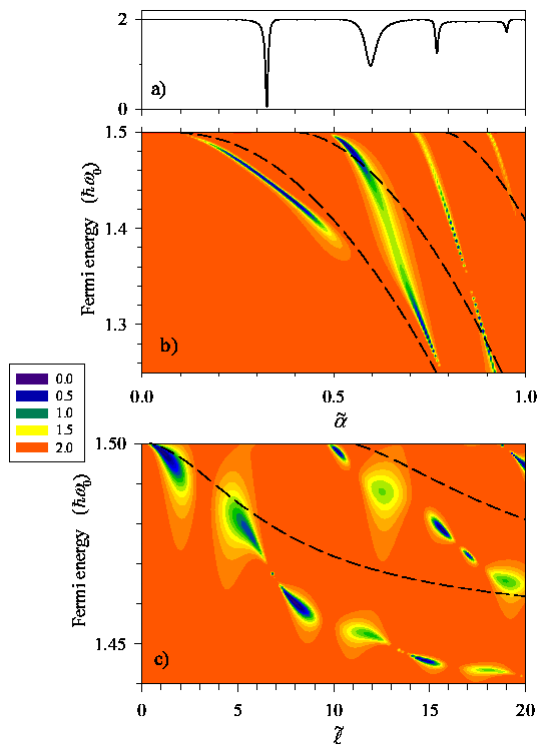,angle=0,width=0.40\textwidth,clip}
\caption{(Color online)
Linear conductance for energies close to the onset of the second plateau
$\varepsilon_2$.
(a) Conductance as a function of the Rashba strength $\tilde\alpha$
for a fixed Fermi energy $E=1.45\hbar\omega_0$ ($\tilde\ell=8$).
(b) Dependence of the conductance with $\tilde\alpha$ for a fixed
dot size $\tilde\ell=8$.
(c) Conductance as a function of the dot
size for $\tilde\alpha=0.3$. Dashed lines
show the 1D bound state energies relative to $\varepsilon_2$.}
\label{fig2}
\end{figure}

In a quantum wire, an initial 2D gas is further confined in the $y$
direction yielding a quasi-1D system with electron propagation
in the longitudinal ($x$) direction for each transverse mode.
Therefore, one would
expect that in the presence of a local Rashba field
the traveling waves be coupled to the bound state due to the
$\{\alpha,p_y\}\sigma_x$ term in Eq.~(\ref{eq_hr}).
The coupling implies opposite spin directions in adjacent subbands
for a parabolic confinement potential \cite{mir01,gov02}.
Since we then have nonresonant
transmission interfering with a channel that passes through
the bound state, a Fano lineshape will develop \cite{zhan05}.
A typical dip in the conductance plateau is shown in Fig.\ \ref{fig1}.
We emphasize that the Rashba interaction given by Eq.~(\ref{eq_hr}) 
plays the role of both the attractive potential
{\em and} the coupling to the continuum states generating Fano
resonances.We refer to this as the {\em Fano-Rashba effect}.
Remarkably, tuning of the Rashba strength would sweep the resonances
across the Fermi energy, see Fig.~\ref{fig2}(a),
much like a gate voltage does in a quantum dot.
Below, we perform exact numerical calculations
to elucidate this effect and put forward an approximate model showing
that the Rashba resonance we find is of a generalized
Fano form \cite{fano}.

We consider a ballistic quasi-1D conductor uniform in the
$x$ direction ,
\begin{equation}\label{eq_h}
H=\frac{p_x^2+p_y^2}{2m} + U(y) + H_R \;,
\end{equation}
where $U(y)$ is the transverse confining potential. It is a good
approximation to take $U(y)=m\omega_0^2 y^2/2$, giving
transverse modes $\varepsilon_n=(n-1/2)\hbar\omega_0$
($n=1,2\ldots$).
The Rashba interaction is localized in a finite region of
the wire and tends to zero for $x\to\pm\infty$.
For definiteness, we assume that $\alpha(x)$ is mostly constant
and equal to $\alpha_0$ in a region of order $\ell$ (a 1D
{\em Rashba dot}) and quickly vanishes outside with a Fermi function
parametrization \cite{Fermi}.
In the results we present below, we take $\hbar\omega_0$ and
$\ell_0=\sqrt{\hbar/m\omega_0}$ as the energy and length units
and use normalized Rashba strength,
$\tilde{\alpha}=\alpha_0/\hbar\omega_0\ell_0$,
and dot length, $\tilde{\ell}=\ell/\ell_0$.

The scattering problem for incoming plane waves is solved
exactly only numerically. We use the quantum transmitting
boundary method \cite{len90} which yields reliable results for
the transmission through nanostructure devices with arbitrary
potentials. Figure~\ref{fig1} shows the results for the linear
conductance as a function of the Fermi energy,
$E$. In the absence of Rashba interaction
the conductance is quantized in steps of $2e^2/h$ every time
$E$ crosses a transverse mode energy threshold.
We also show the conductance in the case of a strong
Rashba interaction $\tilde{\alpha}=1$ for a small dot.
We first note a pronounced dip at energies close to the
onset of the adjacent plateau. The plateaus do not reach
the conductance quanta because the width of the dip is large.
For smaller $\tilde{\alpha}$ the plateaus are not so distorted
but the dip is still apparent, even down to zero conductance,
see Fig. \ref{fig2}(b). We find that
the appearance of dips in the conductance plateaus is quite
general and the particular dip shape depends on the
parameters $\tilde{\alpha}$ and $\tilde\ell$.

The occurrence of dips in the conductance curves of
quantum point contacts \cite{fai90} has been generally
associated with
the existence of attractive impurities inducing strong
backscattering due to resonant reflection \cite{gur93,noc94,gud05}.
Following Refs.~\cite{gur93,noc94}, we expand $\Psi$
in terms of the mode wavefunctions $\phi_n(y)$:
\begin{equation}\label{eq_psi}
\Psi(x,y)=\sum_{ns} \psi_{ns}(x) \phi_n(y)\chi_{s} \,,
\end{equation}
where $\phi_n(y)$ describes the motion in the $y$ direction,
$[-(\hbar^2/2m) d^2/dy^2 +U(y)]\phi_n(y)=\varepsilon_n \phi_n(y)$.
We recall that $\Psi$ is a spinor and ${s}=\{+,-\}$ labels
the spin eigenstates when the spin quantization axis is
along $y$. To determine the amplitudes
$\psi_{ns}$
we consider an energy lying in the first plateau,
$\varepsilon_1<E<\varepsilon_2$, and assume no significant
contributions of states with $n>2$ to the sum in Eq.~(\ref{eq_psi}).
This two-band model has been used to calculate
the energy spectrum of a quantum wire and is a good approach
even for large spin-orbit couplings \cite{gov02}. 

Substituting Eq.~(\ref{eq_psi}) in the Schr\"odinger equation
and projecting onto the mode basis and the spin eigenfunctions
we find,
\begin{eqnarray}
\!\!\!\!
\left[\frac{p_x^2}{2m} -\frac{\{\alpha,p_x\}}{2\hbar}
-E+\varepsilon_1 \right] \psi_{1+}(x) \!\! &=& \!\!
\frac{i\omega_p\alpha}{\hbar} \psi_{2-}(x) \label{eq_ccm1} \;,\\
\!\!\!\!
\left[\frac{p_x^2}{2m} +\frac{\{\alpha,p_x\}}{2\hbar}
-E+\varepsilon_2 \right] \psi_{2-}(x) \!\! &=& \!\!
-\frac{i\omega_p^*\alpha}{\hbar} \psi_{1+}(x)\label{eq_ccm2}
\;,
\end{eqnarray}
which are coupled-channel (CC) equations~\cite{gur93,noc94}.
In Eqs.~(\ref{eq_ccm1},\ref{eq_ccm2}),
$\omega_p=\int{dy\phi_1^* p_y \phi_2}$.
Notably, $\psi_{1+}$ is only coupled to $\psi_{2-}$
because the Rashba intersubband coupling term,
$\{\alpha,p_y\}\sigma_x/2\hbar$,
connects states with opposite spins.
The CC equations for $\psi_{1-}$ and $\psi_{2+}$
are obtained by the replacements $p_x\to-p_x$ and $p_y\to-p_y$. 
Next we introduce the gauge transformation, 
$\psi_{1+,2-}\to\psi_{1+,2-}\exp{(\pm i \int^x{k_R(x') dx')}}$,
and obtain
\begin{eqnarray}
\left[\frac{p_x^2}{2m} -\frac{\hbar^2 k_R^2}{2m}
-(E-\varepsilon_1) \right] \psi_{1+}(x) &=&
V_{12} \psi_{2-}(x) \label{eq_ccm1t} \,,\\
\left[\frac{p_x^2}{2m} -\frac{\hbar^2 k_R^2}{2m}
-(E-\varepsilon_2) \right] \psi_{2-}(x) &=&
V_{21} \psi_{1+}(x)\label{eq_ccm2t}
 \,,
\end{eqnarray}
where we define
\begin{equation}\label{eq_v12}
V_{12}=\frac{i\omega_p \alpha}{\hbar} e^{2i \int^x{k_R(x') dx'}}\,,
\end{equation}
and $V_{21}=V_{12}^*$. Note that we have transformed the original
problem into the CC method of Refs.~\cite{gur93,noc94}. However,
there are two crucial differences. First,
Eqs.~(\ref{eq_ccm1t},\ref{eq_ccm2t}) describe spin states
and, second, the $V_{ij}$ coupling operators have a nontrivial structure
due to the phase factors $\exp{(2i\int^x{k_R(x') dx')}}$.

Equations~(\ref{eq_ccm1t},\ref{eq_ccm2t}) can not,
in general, be exactly
solved with analytical methods. However, a useful approximation utilizes
the ansatz $\psi_{2-}(x)= A \phi_0(x)$ \cite{noc94},
where $A$ is a constant to be found and
$\phi_0(x)$ corresponds to the wavefunction of the 1D bound state; i.e.,
$\phi_0(x)$ satisfies
$(p_x^2/2m-\hbar^2 k_R^2/2m)\phi_0(x)=\varepsilon_0\phi_0(x)$
with $\varepsilon_0<0$.
Since $E>\varepsilon_1$, Eq.~(\ref{eq_ccm1t}) describes a 
1D scattering process with a source term given by 
$V_{12}\psi_{2-}$ and asymptotic wave vector 
$k=\sqrt{2m(E-\varepsilon_1)/\hbar^2}$.
Using the Green function $G$, a formal solution \cite{Joa_book} reads
$|\psi_{1+}\rangle=|\varphi_r\rangle+GV_{12} A |\phi_0\rangle$,
where $|\varphi_r\rangle$ is a solution of the homogeneous
equation behaving for $x\to\infty$
as $te^{ikx}$. As is well known, the Green function for a 1D problem 
is given in terms of $\varphi_r$ and $\varphi_l$, the latter going
as $te^{-ikx}$ for $x\to-\infty$.
These just constitute the background or nonresonant channel
for direct transmission along the waveguide.
Upon substitution in Eq.~(\ref{eq_ccm2t}) we find the total
transmission,
\begin{equation}\label{eq_t}
T_+(E)=|t|^2\frac{(E-\varepsilon_2-\tilde{\varepsilon}_0+\delta)^2
+(\gamma-\Gamma)^2}{(E-\varepsilon_2-\tilde{\varepsilon}_0)^2+\Gamma^2}
\; ,
\end{equation}
where the parameters are determined from 
\begin{eqnarray}
\label{dg1}
\langle\phi_0 |V_{21}GV_{12}|\phi_0 \rangle & \equiv & \Delta+i\Gamma\; ,\\
\label{dg2}
\frac{m}{i\hbar^2kt}\langle\varphi_l^*|V_{12}|\phi_0\rangle
\langle\phi_0|V_{21}|\varphi_r\rangle &\equiv& \delta+i\gamma\; .
\end{eqnarray}
These are complicated functions
of $E$ but we have checked that their variation is slow
around the energy $\varepsilon_2-\varepsilon_0$. Moreover,
the original bound state energy becomes {\em renormalized} to
$\tilde{\varepsilon}_0=\varepsilon_0+\Delta$,
with a shift given by Eq.\ (\ref{dg1}).
We stress that Eq.~(\ref{eq_t}) shows resonance-like
behavior solely due to the intersubband mixing of the Rashba
interaction. The importance of Rashba intersubband mixing
has been emphasized in a number of works
\cite{mir01,gov02,ser05}.
If one neglects it in Eq.~(\ref{eq_hr}),
the bound state $|\psi_{2-}\rangle$
splits off the continuum and does not contribute
to the current. Then, one simply has $T_+(E)=|t|^2$.

The transmission~(\ref{eq_t}) can be cast in the generalized
Fano form,
\begin{equation}
T_+(E)=|t|^2\frac{|\epsilon+q|^2}
{\epsilon^2+1}
\,,
\end{equation}
where $\epsilon=(E-\varepsilon_2-\tilde{\varepsilon}_0)/\Gamma$ 
contains the explicit energy dependence
and the asymmetry parameter $q=\delta/\Gamma+i(\gamma/\Gamma-1)$
is a \emph{complex} quantity. As a result, there is no transmission
zero unless $\gamma=\Gamma$
for some particular set of parameters, see Fig.~\ref{fig2}.
This is a consequence of the nontrivial form of the coupling
operators in Eq.~(\ref{eq_v12}) since for real couplings
(or couplings satisfying special properties \cite{noc94}) one
always finds real $q$'s. The fact that $q$ develops an imaginary part 
has been related to the lack of time-reversal symmetry \cite{kob02,cle01}
but since the Rashba interaction is time-reversal invariant, here we have 
proved that breaking this symmetry is not a necessary condition for
complex $q$'s.

The conductance follows from $\mathcal{G}=(T_{+}+T_{-})e^2/h$,
where the spin-down transmission $T_-$ is found in a similar way to $T_+$ 
solving the CCM equations for $\psi_{1-,2+}$.
We find $T_+=T_-$ since the Rashba interaction
is not able to produce spin polarized currents
due to time-reversal invariance \cite{mol01}.
The result for $\mathcal{G}$ is shown
in Fig.~\ref{fig1}, in rather good agreement
with the exact numerical calculation.

The limit of strong spin-orbit coupling is not, however, realistic.
For an InAs wire with $\alpha\simeq 10$~meV~nm \cite{nit97,sat99,scha04}
and energy spacing $\hbar\omega_0\simeq 1$~meV,
we obtain $\tilde\alpha\simeq 0.2$.
For small $\tilde\alpha$ the state is 
weakly bound and strongly couples to the 
continuum, breaking the validity of
the ansatz for $|\psi_{2-}\rangle$ and
the agreement with the numerics is thus only qualitative.
Nevertheless, the resonance lineshape is still of the Fano
form. In Fig.~\ref{fig2}(b) and (c) we show an exhaustive study of
the dip depth as a function of the Rashba strength and
the dot size. We plot at the same time the position of
the bound states (dashed lines), which do not match
the dip minima due to renormalization. Importantly,
we observe dips going
all the way down to zero within numerical precision.
Increasing $\tilde\alpha$ or $\tilde\ell$ leads to more
bound states. However, the coupling to the continuum
decreases when increasing the level binding. Thus, typically,
two dips coexist in a given plateau provided they are small.
In Fig.~\ref{fig2}(a) we show how, for a fixed value
of $E$, one can tune the resonances
by varying $\tilde\alpha$, which may be regarded
as a gate voltage. This is a central prediction of our work,
which can be tested experimentally.

\begin{figure}
\centerline{
\epsfig{file=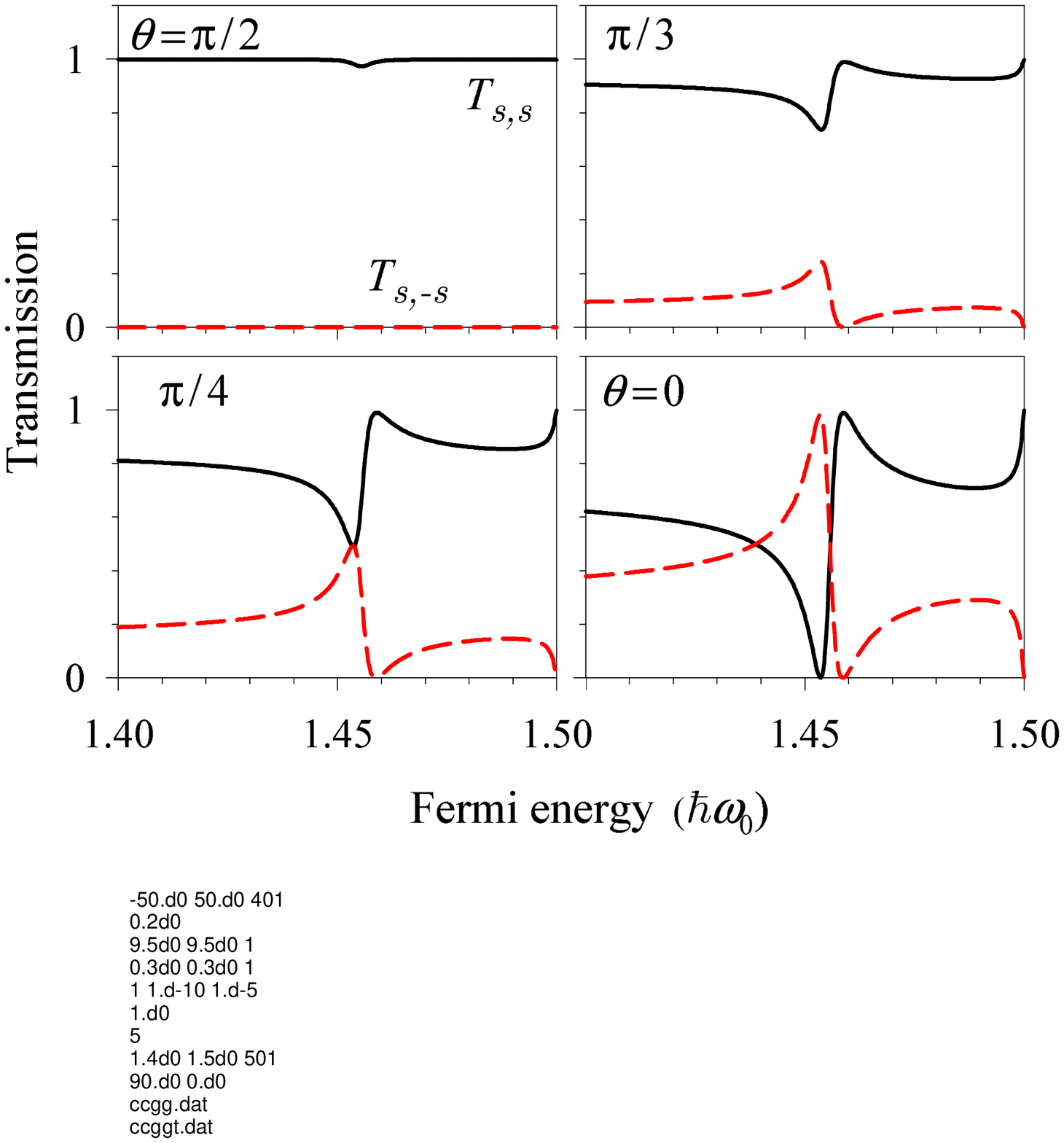,angle=0,width=0.40\textwidth,clip}
}
\caption{Spin dependent transmissions for
$\tilde\alpha=0.3$ and $\tilde\ell=9.5$ when the spin quantization
rotates from $y$ ($\theta=\pi/2$) to $x$ ($\theta=0$).
The total transmission $T=\sum_{s,s'}T_{s,s'}$ is the same in all cases.}
\label{fig3}
\end{figure}

Thus far we have taken the spin quantization axis along $y$,
for which both spin channels are equivalent and uncoupled, i.e.,
$T_+=T_-$ and no spin flip. This is no longer
true along other orientations and, generally,
one must consider the four contributions 
to the total transmission $T=\sum_{s,s'}T_{s,s'}$, where $s$ denotes
the spin state in the left lead (incident)
and $s'$ that of the right lead (transmitted).
Obviously $T$ does not depend on the 
chosen quantization axis. Besides,  
due to time reversal symmetry we always have $T_{++}=T_{--}$
and $T_{+-}=T_{-+}$ \cite{mol01}.
In Fig.~\ref{fig3} we show the spin resolved transmissions
when the spin axis is rotated on the plane from
$\theta=\pi/2$ ($y$-axis) down to 0 ($x$-axis).
Quite remarkably, $T_{s,s}$ and $T_{s,-s}$ develop 
conspicuous Fano lineshapes with exact transmission zeros 
for $T_{s,-s}$ when decreasing $\theta$, even 
for cases in which the total transmission has
a vanishingly small dip. Therefore, experiments using ferromagnetic
leads with tilted spin orientation would greatly enhance
the Fano lineshapes. The precise lineshapes in Fig.\ \ref{fig3} can be explained
with the dependences    
$T_{++}\propto \cos^2(\vartheta/2)$ and
$T_{+-}\propto \sin^2(\vartheta/2)$,
where $\vartheta$ gives 
the phase difference between the transmission amplitudes with
the spins quantized along $y$.
Since $\vartheta$ must change rapidly between 0 and $\pi$
in the presence of a bound state \cite{Joa_book}, there must be a sudden 
variation of the line profile related to the 
scattering through a resonance.

In conclusion, we have demonstrated that for a ballistic
quantum wire
with local spin-orbit coupling the conductance shows
dips that arise from the interference between bound states
formed by the Rashba interaction and the electrons in
the conduction channel. The Rashba interaction is also
responsible for the coupling between the resonant 
and the nonresonant channels, containing phases that change the position
and the asymmetry of the resonance in a nontrivial way.
For strong enough spin-orbit couplings these properties will be 
robust against interaction and charging effects, although 
their detailed investigation requires alternative 
approaches to the scattering theory used here. 
We believe that the observation of the Fano-Rashba effect addressed in this work 
is within the scope of present experimental techniques.

\acknowledgments

We acknowledge R. L\'opez for valuable discussions.
This work was supported by the Grant No.\ FIS2005-02796
(MEC) and the ``Ram\'on y Cajal'' program.


\begin{thebibliography}{90}
\bibitem{Fermi}
We take $\alpha(x)=\alpha_0 [f(x-\ell/2)-f(x+\ell/2)]$, with 
$f(x)=1/(1+e^{x/\sigma})$.
In the numerical simulations we assume a small diffusivity $\sigma=0.2\ell_0$
but the results are not very sensitive to the precise value of $\sigma$
provided $\sigma<\ell$.
\end{thebibliography}
\end{document}